\pdfoutput=1

\documentclass[11pt,twoside,a4paper,cmspaper,final,collab]{cms-tdr}
\def\svnVersion{404335}\def\cmsCernNoTag{CERN-EP/2016-326}\def\cmsCernDate{\today}\def\cmsMessage{Published in the Journal of High Energy Physics as \href{http://dx.doi.org/10.1007/JHEP05(2017)029}{\doi{10.1007/JHEP05(2017)029.}}}

\begin{document}\cmsNoteHeader{B2G-16-001}

\hyphenation{had-ron-i-za-tion}
\hyphenation{cal-or-i-me-ter}
\hyphenation{de-vices}
\RCS$Revision: 395477 $
\RCS$HeadURL: svn+ssh://svn.cern.ch/reps/tdr2/papers/B2G-16-001/trunk/B2G-16-001.tex $
\RCS$Id: B2G-16-001.tex 395477 2017-03-22 09:54:42Z aspiezia $
\newlength\cmsFigWidth
\ifthenelse{\boolean{cms@external}}{\setlength\cmsFigWidth{0.85\columnwidth}}{\setlength\cmsFigWidth{0.4\textwidth}}
\ifthenelse{\boolean{cms@external}}{\providecommand{\cmsLeft}{top\xspace}}{\providecommand{\cmsLeft}{left\xspace}}
\ifthenelse{\boolean{cms@external}}{\providecommand{\cmsRight}{bottom\xspace}}{\providecommand{\cmsRight}{right\xspace}}

\providecommand{\PQT}{\ensuremath{\mathrm{T}}\xspace}
\providecommand{\POWHEGBOX}{\textsc{powheg box}\xspace}
\cmsNoteHeader{B2G-16-001}
\title{Search for single production of vector-like quarks decaying to a Z boson and a top or a bottom quark in proton-proton collisions at $\sqrt{s} = 13$\TeV}

\date{\today}

\abstract{
A search for single production of vector-like quarks, T and B, decaying into a Z boson and a top or a bottom quark, respectively, is presented. The search is performed using data collected by the CMS experiment at the LHC in proton-proton collisions at $\sqrt{s}=13$\TeV, corresponding to an integrated luminosity of 2.3\fbinv. An exotic T quark production mode through the decay of a heavy $\PZpr$ resonance is also considered. The search is performed in events with a Z boson decaying leptonically, accompanied by a bottom or a top quark decaying hadronically. No excess of events is observed over the standard model background expectation.
Products of production cross section and branching fraction for T and B quarks from 1.26 and 0.13\unit{pb} are excluded at 95\% confidence level for the range of resonance mass considered, which is between 0.7 and 1.7\TeV.
Limits on the product of the \PZpr boson production cross section and branching fraction, with the \PZpr boson decaying to the Tt final state, are set between 0.31 and 0.13\unit{pb}, for \PZpr boson masses in the range from 1.5 to 2.5\TeV.
This is the first search at 13\TeV for single production of vector-like quarks in events with a Z boson decaying leptonically accompanied by boosted jets.
}

\hypersetup{%
pdfauthor={CMS Collaboration},%
pdftitle={Search for single production of vector-like quarks decaying to a Z boson and a top or a bottom quark in proton-proton collisions at sqrt(s) = 13 TeV},%
pdfsubject={CMS},%
pdfkeywords={CMS, physics, B2G, heavy quarks}}

\maketitle
\section{Introduction}
The discovery of a Higgs boson~\cite{Aad:2012tfa, Chatrchyan:2012xdj, Chatrchyan:2013lba} with a mass of 125\GeV~\cite{Aad:2015zhl} by the ATLAS and CMS experiments confirmed the success of the standard model (SM) in predicting a wide range of high-energy phenomena. However, several questions related to the nature of electroweak symmetry breaking
remain unanswered and, to address them, several new theoretical models have been proposed such as little Higgs~\cite{Schmaltz:2005ky}, large extra dimensions~\cite{Antoniadis:2001cv, Randall:1999ee}, and composite Higgs models~\cite{Agashe:2004rs}. Many of these models predict the existence of heavy resonances with masses of the order of 1\TeV, called vector-like quarks (VLQs)~\cite{Aguilar-Saavedra:2013qpa, AguilarSaavedra:2009es, DeSimone:2012fs, Matsedonskyi:2014mna, Buchkremer:2013bha}. These are hypothetical new spin-1/2 particles with the property that left- and right-handed (LH and RH) chiralities transform in the same way under the SM symmetry group $SU(2)_L \times U(1)_Y \times SU(3)_C$. As a consequence, they do not receive mass through a Yukawa coupling term, as do the chiral fermions of the SM, but through a direct mass term of the form $m\overline{\psi}\psi$. A fourth generation of chiral quarks is strongly disfavoured by the precision SM measurements~\cite{Eberhardt:2012sb}, because of the modifications that the Yukawa term would bring to the Higgs production cross section and branching fractions ($\mathcal{B}$s). The VLQs are not similarly constrained. Previous searches for VLQs have been performed by both ATLAS~\cite{Aad:2015kqa, Aad:2015mba, Aad:2014efa, Aad:2016shx, Aad:2016qpo, Aad:2015voa} and CMS~\cite{Khachatryan:2015oba, Khachatryan:2015gza, Chatrchyan:2013wfa, Khachatryan:2016vph}, using data samples collected at $\sqrt{s}= 7$, 8, and 13\TeV. This paper presents the first search at 13\TeV for single production of VLQs in final states with boosted jets and a leptonically decaying Z boson.

\begin{figure}[!b]
\centering
\includegraphics[scale=0.34]{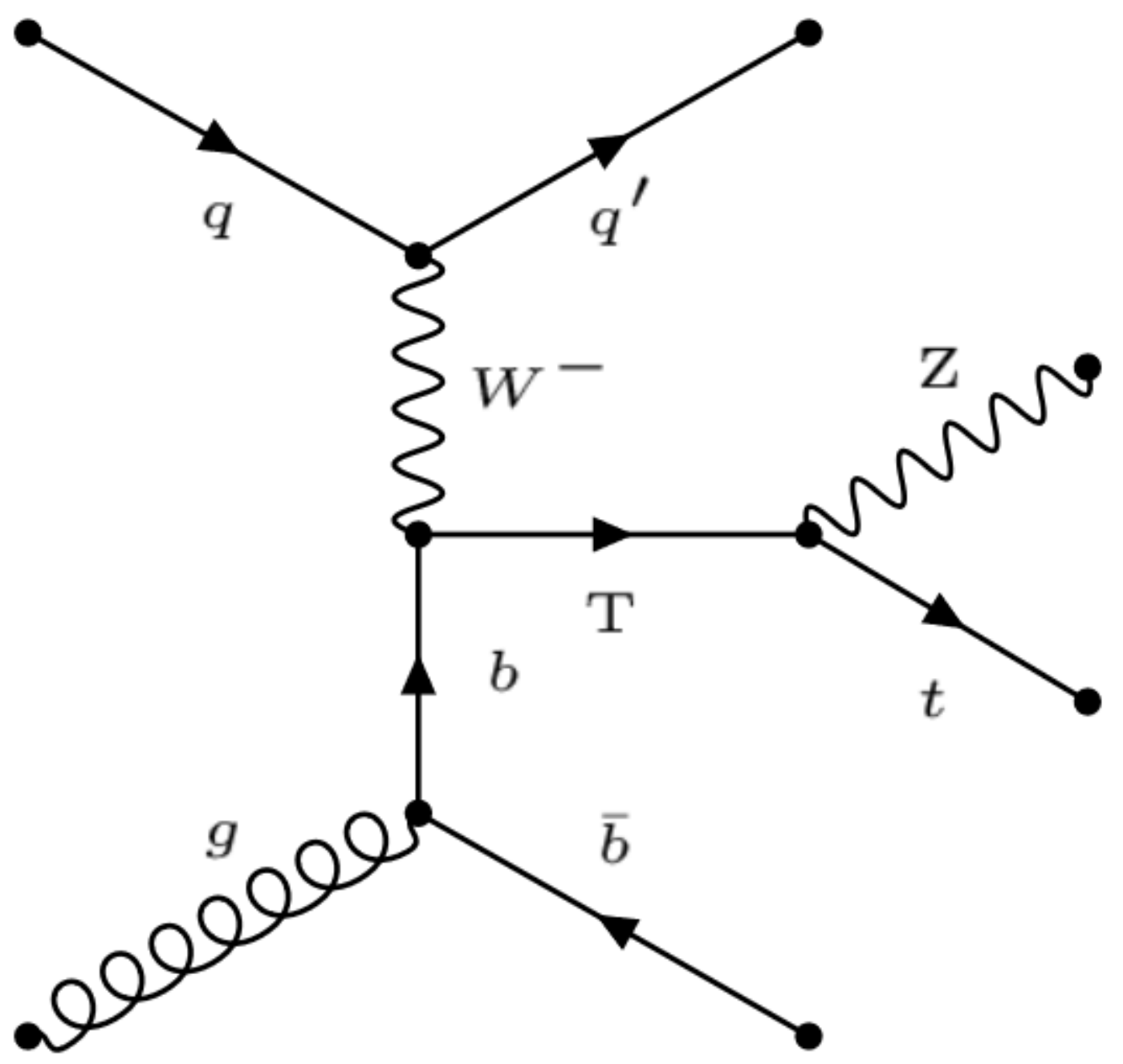}
\includegraphics[scale=0.34]{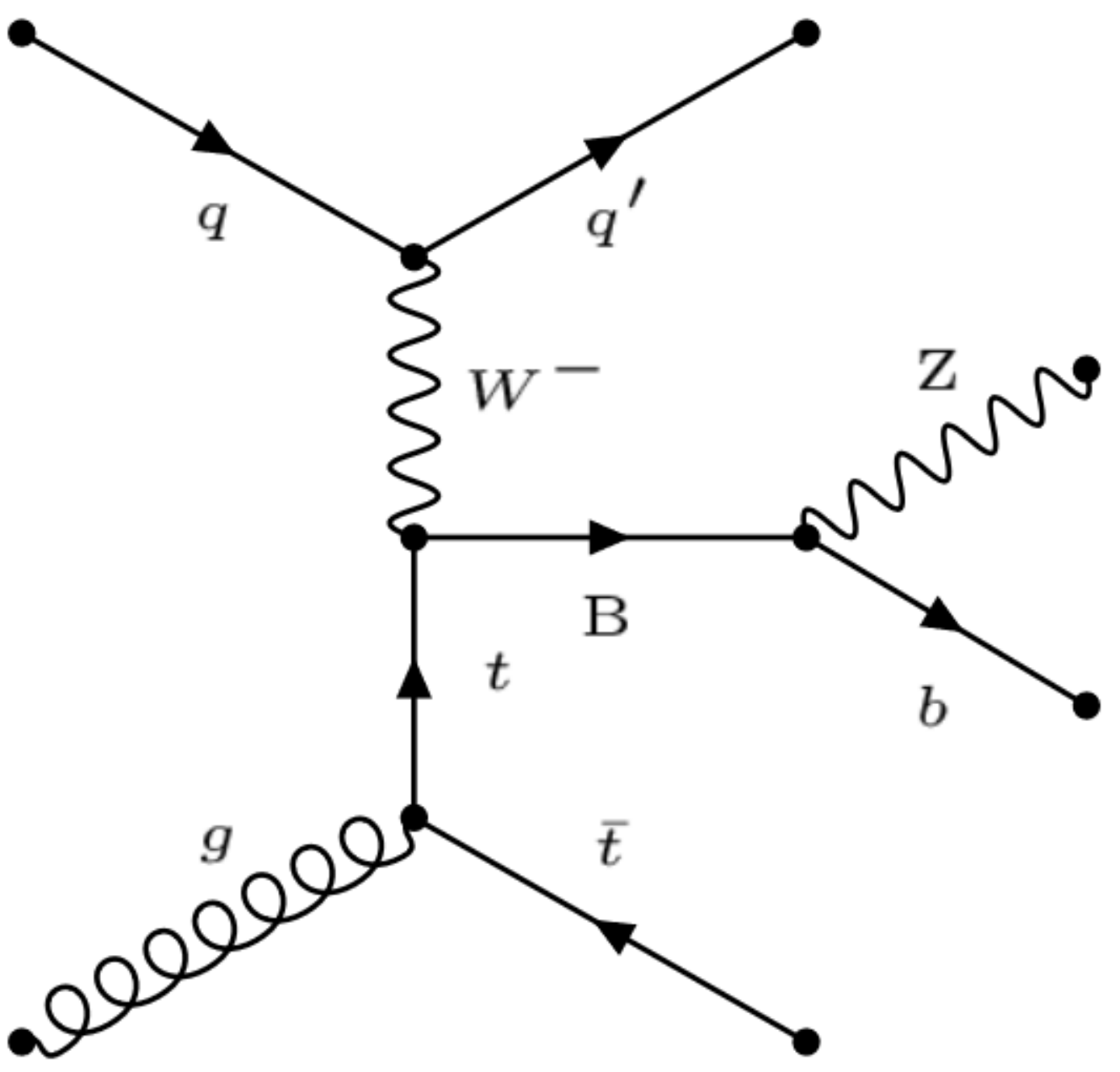}
\includegraphics[scale=0.29]{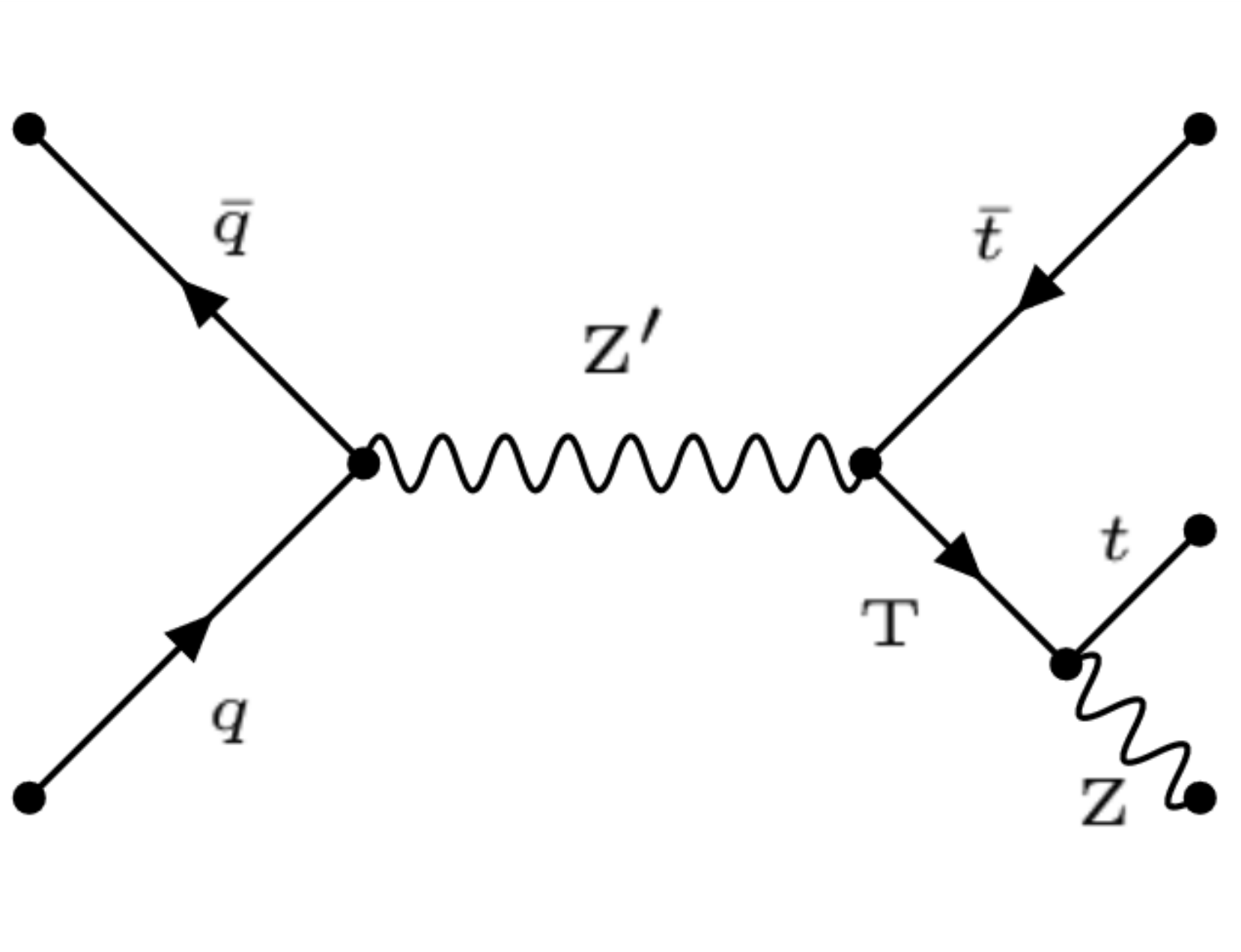}
\caption{Leading order Feynman diagrams for the production of a single T (B) vector-like quark and its decay to a Z boson and a t (b) quark on the left (center) and production of a \PZpr boson decaying to Tt on the right.}\label{fig:feynman}

\end{figure}

In a model-independent approach~\cite{Aguilar-Saavedra:2013qpa}, VLQs can be grouped in multiplets (singlet, doublet, triplet, etc.) that couple to the SM particles. Singlets include the T and B quarks with charges of $+2/3$ and $-1/3$, respectively. Multiplets include the T and B quarks and exotically charged VLQs labelled X and Y, which have charges of $+5/3$ and $-4/3$, respectively. In this analysis, we present a search for a singlet or a doublet T quark that decays to a Z boson and a t quark, with subsequent decays Z~$\to\ell^+\ell^-$, where $\ell$ can be a muon or an electron, and $\PQt\to\PQb\PW\to\PQb\PQq\PQq'$. We also present the search for a singlet B quark, decaying to a b quark and a Z boson that decays to $\ell^+\ell^-$. Examples of Feynman diagrams for the single production of T and B VLQs are shown in Fig.~\ref{fig:feynman}.

A singlet T quark has three different decay channels into SM particles: bW, tZ, and tH~\cite{Aguilar-Saavedra:2013qpa, AguilarSaavedra:2009es, DeSimone:2012fs, Matsedonskyi:2014mna, Buchkremer:2013bha}. Using the equivalence theorem~\cite{Lee:1977eg} the branching fractions for these three decay modes are 0.5, 0.25, and 0.25, respectively. The T doublet can decay to tZ or tH, each with a branching fraction of 0.5. In this case the doublet structure of the fermion multiplet produces a tree level coupling of the T to the neutral bosons, but not to the W. As we are neglecting possible mass mixings between the T and the t quarks, the branching fraction of the doublet T to the tW final state is taken to be zero. Similarly, the decay modes for a B singlet are tW (branching fraction of 0.5), bZ (0.25), and bH (0.25).
The couplings of the new particles to SM particles can be described with the following coefficients: $C(\PQb\PW)$ (for the $\Pp\Pp\to\PQT\PQb$ process), $C(\PQt\PW)$ (for the B(t) final state), $C(\PQt\Z)$ (for T(t)), and $C(\PQb\Z)$ (for B(b)). For singlet VLQs, the RH chiralities are suppressed compared to the LH ones by a factor proportional to the standard quark mass over the VLQ mass, while for doublets it is the LH chirality that is suppressed~\cite{Cacciapaglia:2011fx}.

An additional production mode is also investigated for the T quark, i.e. the production of a neutral spin-1 heavy \PZpr boson~\cite{Bini:2011zb, Accomando:2010fz, Greco:2014aza} that decays to a Tt final state, as shown by the Feynman diagram in Fig.~\ref{fig:feynman}.

The mass range for the T and B quarks studied in this analysis is 0.7--1.7\TeV, while the \PZpr boson is searched for in the 1.5--2.5\TeV range. Lower masses of VLQs are not investigated because pair production searches of VLQs have excluded masses below 0.7--0.9\TeV~\cite{Aad:2015kqa, Aad:2015mba, Aad:2014efa, Aad:2016qpo, Khachatryan:2015oba, Khachatryan:2015gza}. Furthermore, at high masses single production modes are favoured over the pair production modes. In the mass ranges considered in the analysis, the t quark from the T quark decay can be produced with high transverse momentum (\pt), resulting in a final state where the decay products of the t quark are emitted close to each other in a topology with overlapping and merged jets. For this reason, final states with t quark jets (t jets) and W boson jets (W jets) are investigated, \ie in events with large-cone jets that are identified using jet grooming techniques~\cite{Khachatryan:2014gha, Khachatryan:2015axa} as coming from the hadronic decay of a t quark or a W boson. Jet grooming techniques are used to reduce the impact of the underlying event and the presence of additional primary vertices in the events (pileup), and of low \pt gluon radiation, i.e. particles that are not related to the hard process. Evidence for the production of new particles is searched for in the reconstructed candidate heavy quark mass spectrum.

\section{The CMS detector, data sample and simulation}\label{sec:CMS}
The general-purpose CMS detector operates at one of the four interaction points of the LHC. Its central feature is a 3.8\unit{T} superconducting solenoid magnet with an internal diameter of 6 m. The following subdetectors are found within the magnet volume: the silicon tracker, the crystal electromagnetic calorimeter (ECAL), and the brass and scintillator hadron calorimeter (HCAL). Muons are measured in gas-ionization detectors embedded in the steel flux-return yoke outside the solenoid. In addition, the CMS detector has extensive forward calorimetry: two steel and quartz-fiber hadron forward calorimeters, which extend the coverage to regions close to the beam pipe.
A more detailed description of the CMS detector, together with a definition of the coordinate system used and the relevant kinematic variables, can be found in Ref.~\cite{Chatrchyan:2008aa}.

The analysis is based on the data sample collected by the CMS experiment in proton-proton collisions at a center-of-mass energy of 13\TeV in 2015, corresponding to an integrated luminosity of 2.3\fbinv. Events with a Z boson decaying to muons are preselected by a single-muon trigger, requiring the presence of an isolated muon with $\pt>20$\GeV. Events with the Z boson decaying to electrons are preselected with a single-electron trigger that requires the presence of an electron with $\pt>105$\GeV. This high \pt threshold does not degrade the signal efficiency, since the electrons of interest would come from the decay chain of a high mass resonance.

Background samples are generated using \MADGRAPH~5.2~\cite{Alwall:2014hca} for Z/$\gamma$*+jets, \ttbar+V, and tZq processes and {\POWHEGBOX v2}~\cite{Nason:2004rx, Frixione:2007vw, Alioli:2010xd, Alioli:2011as} for \ttbar and single t quark production, interfaced to \PYTHIA 8.212 \cite{Sjostrand:2014zea}, which uses tune {CUETP8M1}~\cite{Khachatryan:2015pea} for the description of hadronisation and fragmentation. The standalone \PYTHIA generator is used to simulate SM diboson production.

Signal samples are generated using \MADGRAPH~5.2 interfaced with \PYTHIA, for T and B quark masses between 0.7 and 1.7\TeV in steps of 0.1\TeV, and for three \PZpr mass hypotheses: 1.5, 2.0, and 2.5\TeV. Singlet and doublet T quarks and singlet B quarks, with both LH and RH couplings to the SM particles, are simulated.
Theoretical cross sections used in the analysis are reported in Table~\ref{tab:CrossSec} as calculated in Ref.~\cite{Matsedonskyi:2014mna}, where a simplified approach is used to allow model-independent interpretation of the experimental results. The theoretical width of the VLQs is negligible compared to the experimental mass resolution, for values of the couplings $C(\PQb\PW)$, $C(\PQt\PW)$, $C(\PQt\Z)$, and $C(\PQb\Z)$, equal to or below 0.5.

\begin{table}[!t]
\centering
\topcaption{Theoretical cross sections for T(b), B(t), B(b), and T(t) processes for the different benchmark mass points considered in the analysis, with the couplings set to 0.5 as calculated at NLO in Ref.~\cite{Matsedonskyi:2014mna}. Cross sections do not depend on the chirality of the new particle (T or B).}\label{tab:CrossSec}
\begin{tabular}{ccccc}
\hline
Mass [\TeVns{}] & $\sigma(\Pp\Pp\to\PQT\PQb)$\,[pb] & $\sigma(\Pp\Pp\to\PB\PQt)$\,[pb]  & $\sigma(\Pp\Pp\to\PB\PQb)$\,[pb]  & $\sigma(\Pp\Pp\to\PQT\PQt)$\,[pb] \\
\hline
0.7 & 1.455 & 0.186 & 1.085 & 0.125 \\
0.8 & 0.965 & 0.133 & 0.754 & 0.091 \\
0.9 & 0.680 & 0.097 & 0.555 & 0.068 \\
1.0 & 0.488 & 0.071 & 0.413 & 0.051 \\
1.1 & 0.338 & 0.053 & 0.298 & 0.038 \\
1.2 & 0.246 & 0.040 & 0.224 & 0.029 \\
1.3 & 0.179 & 0.030 & 0.170 & 0.022 \\
1.4 & 0.135 & 0.023 & 0.132 & 0.017 \\
1.5 & 0.102 & 0.018 & 0.104 & 0.014 \\
1.6 & 0.076 & 0.014 & 0.080 & 0.011 \\
1.7 & 0.058 & 0.011 & 0.062 & 0.008 \\
\hline
\end{tabular}

\end{table}

The generated events are passed through a simulation of the CMS detector using \GEANTfour~\cite{Agostinelli:2002hh, Allison:2006ve}. The pileup distribution in simulation is matched to the observed distribution of additional interactions in data.
Samples are generated with {NNPDF}~3.0~\cite{Ball:2014uwa} parton distribution function sets.

\section{Physics object reconstruction}\label{sec:object}
Primary vertices are reconstructed using a deterministic annealing filter algorithm~\cite{Chatrchyan:2014fea}. The interaction vertex corresponding to the hard scattering is chosen as the one that maximizes
the squared \pt sum of the clustered physics objects associated with it. Selected events are required to have a primary vertex within 24 cm of the mean interaction point in the $z$-direction and within 2 cm in the $x$-$y$ plane.

A particle-flow (PF) algorithm~\cite{CMS-PAS-PFT-09-001, CMS-PAS-PFT-10-001} is used to identify and to reconstruct charged and neutral hadrons, photons, muons, and electrons, through an optimal combination of the information from the entire detector. Muon candidates are reconstructed by combining the information from the silicon tracker and the muon system in a global track fit~\cite{Chatrchyan:2012xi}. Muons are then required to be isolated, to satisfy $\pt>20$\GeV and $\abs{\eta} <$ 2.4, and to pass additional identification criteria based on the track impact parameter, the quality of the track reconstruction, and the number of hits recorded in the tracker and the muon systems. The leading muon is required to have $\pt>22$\GeV, which ensures that selected muons are in a region of high trigger efficiency.

Electron candidates are reconstructed by combining the information from the ECAL and from the silicon tracker~\cite{Khachatryan:2015hwa}. Electrons are then selected if they are isolated and if they have $\pt>20$\GeV and $\abs{\eta} <$ 2.5. Additional requirements are applied on the ECAL shower shape, on the variables related to the track-cluster matching, on the impact parameter, and on the ratio of energies measured in HCAL and ECAL in the region around the electron candidate. The leading electron is required to have $\pt>115$\GeV, i.e. to be in the region of high trigger efficiency.

For both muons and electrons, a lepton isolation variable is used to reduce background coming from events where one hadronic jet is misidentified as a lepton. This variable is defined as the sum of the \pt of the charged hadrons, neutral hadrons, and photons found in a cone, defined as $\Delta R = \sqrt{\smash[b]{(\Delta\eta)^2+(\Delta\phi)^2}}$ (where $\phi$ is azimuthal angle), around the original lepton track, corrected for the effects of pileup~\cite{Chatrchyan:2012xi, Khachatryan:2015hwa}, and divided by the lepton \pt. The cone size used is 0.4 for muons and 0.3 for electrons.

Jet candidates are clustered starting from the PF candidates using the anti-\kt clustering algorithm~\cite{Cacciari:2008gp} with distance parameters of 0.4 (``AK4 jets'') and 0.8 (``AK8 jets''). The jet energy scale (JES) is calibrated through correction factors dependent on the \pt and $\eta$ of the jet. The jet energy resolution (JER) for the simulated jets is smeared in order to reproduce the actual detector resolution observed in data.
Jet candidates are required to have angular separation $\Delta R>0.4\,(0.8)$ from identified leptons for AK4 (AK8) jets, and are selected with $\pt>25$\,(180)\GeV and with $\abs{\eta}< 2.4.$
A pruning algorithm~\cite{Ellis:2009su} is applied to AK8 jets to tag those that originate from the hadronic decay of a W boson.
The mass of the jet, after the pruning is performed, is used as a discriminant to select W bosons and reject quark and gluon jets.

The other variable used to discriminate the W jet from quark and gluon jets is the $N$-subjettiness~\cite{Khachatryan:2014vla}. This is a measure of how consistent a jet is with having N or fewer subjets. This variable is defined as:
\begin{equation}\label{equ:subjettiness}
\tau_N = \frac{1}{d_0} \, \sum_k [ {\pt}_{k} \, \min(\Delta R_{1,k},\Delta R_{2,k},\ldots\Delta R_{N,k})],
\end{equation}
where $k$ is the index ranging over the PF particles that form the jet, ${\pt}_{k}$ is the transverse momentum of the $k$th constituent, $\Delta R_{n,k}$ is the distance between the $k$th constituent and the $n$th subjet axis, $d_0=\sum_kp_{\mathrm{T}k}R_0$ is a normalization factor, with $R_0$ equal to the original jet distance parameter, and $N$ is the number of subjets under consideration. The final variable used to discriminate W jets, which are expected to have two subjets, from quark and gluon jets, which are expected to have no subjets, is $\tau_{21} = \tau_2 / \tau_1$. An AK8 jet is W-tagged if the mass range of the pruned jet is within 65--105\GeV and if $\tau_{21}$ is lower than 0.6. The efficiency of the W-tagging procedure is corrected for discrepancies between data and simulation~\cite{Khachatryan:2014gha, Khachatryan:2014hpa, Khachatryan:2015dcf}. In a similar way, AK8 jets can be identified as coming from the hadronic decay of a t quark. These t quark jets are required to pass the following selections: $\pt>400$\GeV, mass of the jet reconstructed with the modified mass drop tagger algorithm~\cite{Dasgupta:2013ihk, Larkoski:2014wba} between 110 and 210\GeV, and $\tau_{32} = \tau_3 / \tau_2$, defined using Eq.~(\ref{equ:subjettiness}), lower than 0.69. Also in this case, scale factors are applied to correct for disagreement between data and simulation. Finally, AK4 jets can be tagged as coming from a b quark using the combined secondary vertex algorithm~\cite{CMS-PAS-BTV-15-001}. A ``medium'' working point with efficiency of 70\% on real b jets and rejection of 99\% of light-flavor jets is used together with a ``loose'' working point, which has 85\% of efficiency and 90\% of rejection.

\section{Event selection}\label{sec:selection}
In this analysis, we search for a Z boson decaying to leptons, and a t or b quark arising from the decay of a T or B quark, respectively. Events are required to have two muons or electrons forming a Z boson candidate with an invariant mass between 70 and 110\GeV, and are sorted into six categories: four for the T search, and two for the B search. A t quark from a T quark decay can be identified in three different scenarios: fully merged (a t quark jet is identified), partially merged (a W jet and a b jet are identified), or resolved (three AK4 jets are reconstructed). Thus we define four categories of events for the T search:
\begin{itemize}
\item category 0: $\PQT\to 2\ell + \text{1\PQt jet}$;
\item category 1: $\PQT\to 2\ell + \text{1 \PW{} jet} + \text{1 \PQb jet}$;
\item category 2: $\PQT\to 2\mu + \text{1 \PQb jet} +\text{2 jets}$;
\item category 3: $\PQT \to 2\Pe + \text{1 \PQb jet} + \text{2 jets}$
\end{itemize}
where the b jet is tagged with the ``medium'' working point.

\begin{table}[!b]
\centering
\topcaption{Summary of the final event selection for the four categories of the T search. In each category exactly two oppositely charged leptons are required.}\label{tab:Selection1}
\begin{tabular}{lcccc} \hline
               & 2$\ell$ + 1t jet & 2$\ell$ + 1W jet + 1b jet & 2$\mu$ + 1b jet + 2 jets & 2e + 1b jet + 2 jets\\\hline
Leptons        & 2$\mu$/2e        & 2$\mu$/2e                 & 2$\mu$                   & 2e                 \\
Lead lep \pt   & {$>$115\GeV}& {$>$115\GeV}& {$>$115\GeV}& {$>$115\GeV}\\[2ex]
\multirow{2}[0]{*}{Jet}            & \multirow{2}[0]{*}{one t jet} & one W jet & \multicolumn{2}{c}{three AK4 jets } \\
                                   &                               & one b jet & \multicolumn{2}{c}{(one b-tagged)}  \\[2ex]
t \pt                              & $>$400\GeV     & {$>$150\GeV}    & {$>$150\GeV}    & {$>$150\GeV}    \\
$\Delta R(\ell,\ell)$              & $<$1.1         & $<$1.0     & {$<$0.9}  & {$<$0.9} \\
N(b jet) & {$\geq$1} & {$\geq$1}& {$\geq$1}& {$\geq$1}\\\hline
\end{tabular}

\end{table}

\begin{table}[!b]
\centering
\topcaption{Summary of the final event selection for the two categories of the B search. In each category exactly two oppositely charged leptons are required.}\label{tab:Selection2}
\begin{tabular}{lcc}
\hline
                      & 2$\mu$ + 1b jet & 2e + 1b jet    \\\hline
Leptons               & 2$\mu$          & 2e             \\
Lead lep \pt          & \multicolumn{2}{c}{$>$ 115\GeV} \\
Jet                   & \multicolumn{2}{c}{one b jet with $\pt>150$\GeV} \\
$\Delta R(\ell,\ell)$ & \multicolumn{2}{c}{$<$0.7}     \\
N(b jet)              & \multicolumn{2}{c}{$\geq$2}    \\\hline
\end{tabular}

\end{table}

The electron and muon identification efficiencies are different, therefore resolved events with two muons and resolved events with two electrons are considered separately. The fully merged and partially merged topology events, where the Z decays to muons or to electrons, are considered together to increase the numbers of events in the control samples. If one event falls in two or more categories, the first one of these categories is considered. For example if one event falls in both categories 0 and 3, it will be considered only in category 0.
In category 0, the t jet with the highest \pt (and $\pt>400$\GeV) is retained as the t quark candidate. For category 1, the t quark candidate is reconstructed by summing the Lorentz vectors of the W jet and the b jet, while for categories 2 and 3 the sum is made for the three jets. In these last three categories a minimum \pt of 150\GeV is required for the t quark candidate, and if more than one t candidate is found, the one with the invariant mass closest to the t quark mass is selected.

In addition to requiring a Z boson and a t quark in the event, for each category at least one b jet has to be present, the two leptons from the Z boson decay have to be close to each other ($\Delta R< 0.9$--1.1, depending on the category), and the leading lepton (muon or electron) must have $\pt>115$\GeV.

The B quark candidate is reconstructed by combining together the Z boson and the b jet (tagged with the ``medium'' working point) with the highest \pt in the event. Two categories are defined, depending on whether the Z boson decays to muons or electrons. Further requirements applied to reduce the background are: the b jet $\pt>150$\GeV, at least 2 b jets are present in the event, $\Delta R$ between the two leptons is lower than 0.7, and the leading lepton (muon or electron) $\pt>115$\GeV.

After the full event selection, which is summarized respectively for the T and the B searches in Tables~\ref{tab:Selection1} and~\ref{tab:Selection2}, the masses of the T and B quark candidate are reconstructed and required to be above 500\GeV. The following experimental mass resolutions are evaluated from simulation, for four different signal hypotheses: 16\% for T(b), 24\% for T(t), 14\% for B(t), and 12\% B(b). The number of expected signal events and signal efficiencies are shown in Tables~\ref{tab:Estimation1} and~\ref{tab:Estimation2} for the T and the B search respectively.

The background is largely dominated by Z+jets events (between 80\% and 92\%, depending on the category), with smaller contributions from other backgrounds (t$\bar{\mbox{t}}$+V, tZq, \ttbar, single t quark, and SM diboson production).

\begin{table}[!b]
\centering
\topcaption{The numbers of estimated background events compared to the measured numbers of events for the four categories of the T search. The quoted uncertainties in the background estimates include both statistical and systematic components, as described in Section~\ref{sec:systematics}. Expected signal yields and signal efficiencies (in parentheses) are also shown for three benchmark mass points.}\label{tab:Estimation1}
\begin{tabular}{lcccc}
\hline
Channel         & 2$\ell$ + 1t jet & 2$\ell$ + 1W jet + 1b jet & 2$\mu$ + 1b jet + 2 jets & 2e + 1b jet + 2 jets \\\hline
Estimated  background& {1.4 $\pm$ 0.8} &{8.6 $\pm$ 1.7} & {126.5 $\pm$ 14.0} & {75.1 $\pm$ 8.1} \\
 Observation      & 0 & 7 & 109 & 91 \\\hline
T(b) LH, $M=0.7$\TeV & 0.07 (0.1\%) & 1.6  (3\%) &  6.3 (11\%) &  4.3 (8\%)\\
T(b) LH, $M=1.2$\TeV & 0.6  (6\%)   & 0.3  (3\%) &  1.3 (13\%) &  0.9 (9\%)\\
T(b) LH, $M=1.7$\TeV & 0.2  (9\%)   & 0.05 (2\%) &  0.2 (11\%) &  0.2 (8\%)\\\hline
\end{tabular}
\end{table}

\begin{table}[!bht]
\centering
\topcaption{The numbers of estimated background events compared to the measured numbers of events for the two categories of the B search. The quoted uncertainties in the background estimates include both statistical and systematic components, as described in Section~\ref{sec:systematics}. Expected signal yields and signal efficiencies (in parentheses) are also shown for three benchmark mass points.}\label{tab:Estimation2}
\begin{tabular}{lcc}
\hline
Channel     & 2$\mu$ + 1b jet                    & 2e + 1b jet                       \\\hline
Estimated   background& {7.0 $\pm$ 0.8}  & {4.1 $\pm$ 0.5} \\
Observation & 8                                  & 3                                 \\
B(t) LH, $M=0.7$\TeV & 0.4  (5\%) & 0.2  (4\%)\\
B(t) LH, $M=1.2$\TeV & 0.1  (7\%) & 0.09 (6\%)\\
B(t) LH, $M=1.7$\TeV & 0.02 (6\%) & 0.02 (5\%)\\\hline
\end{tabular}

\end{table}

\begin{figure}[!bth]
\centering
\includegraphics[width=0.48\textwidth]{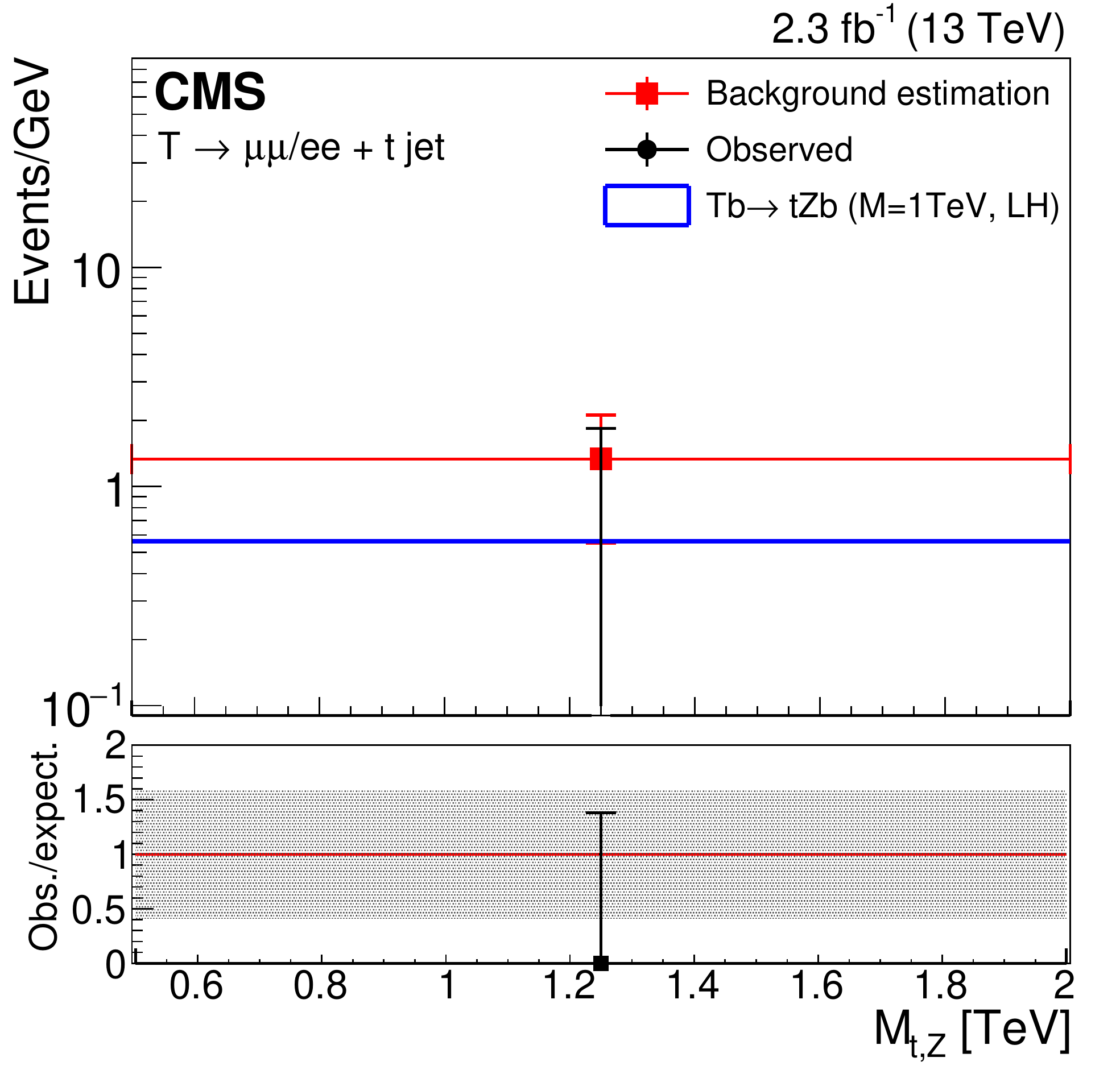}
\includegraphics[width=0.48\textwidth]{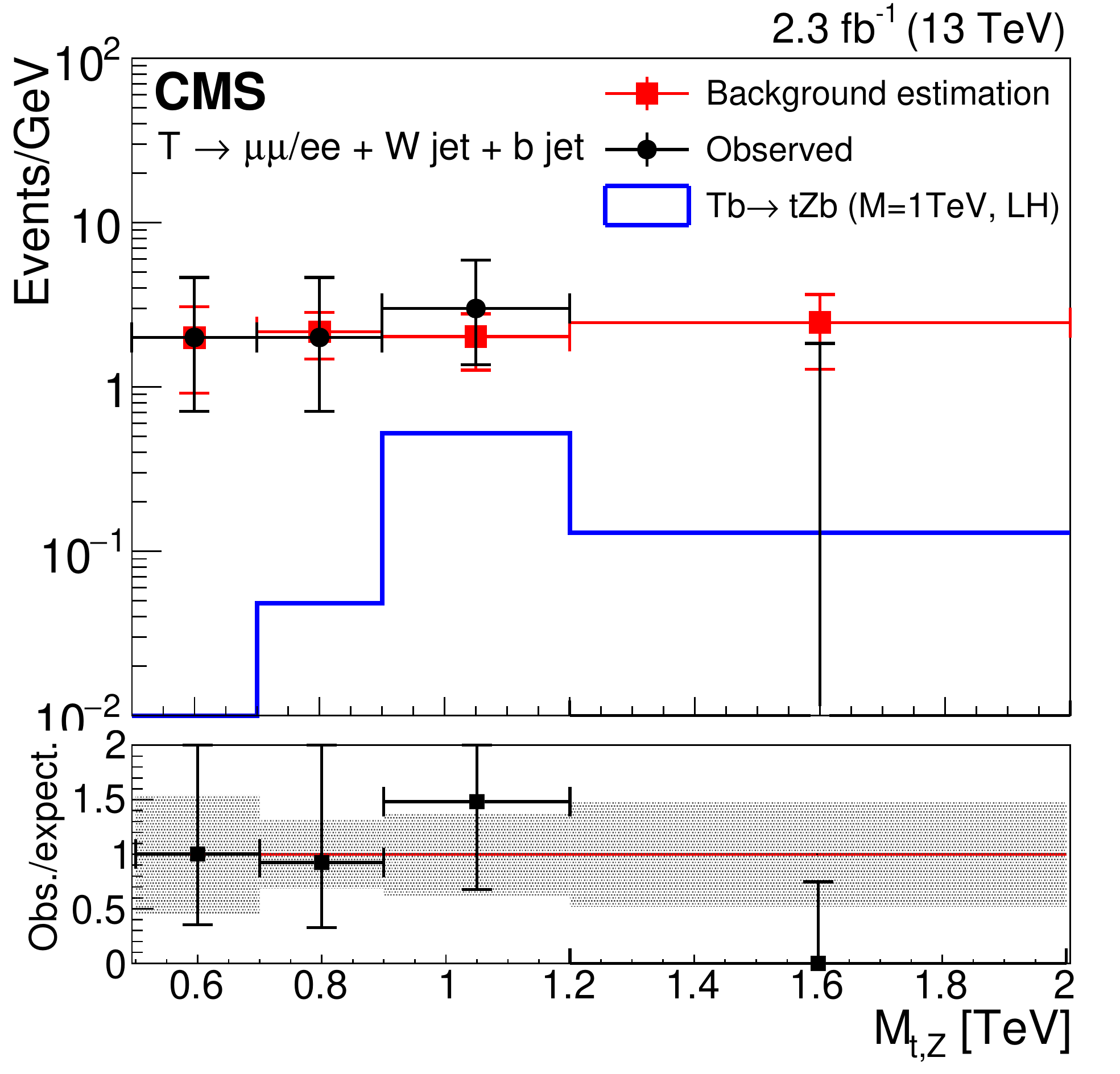}\\
\includegraphics[width=0.48\textwidth]{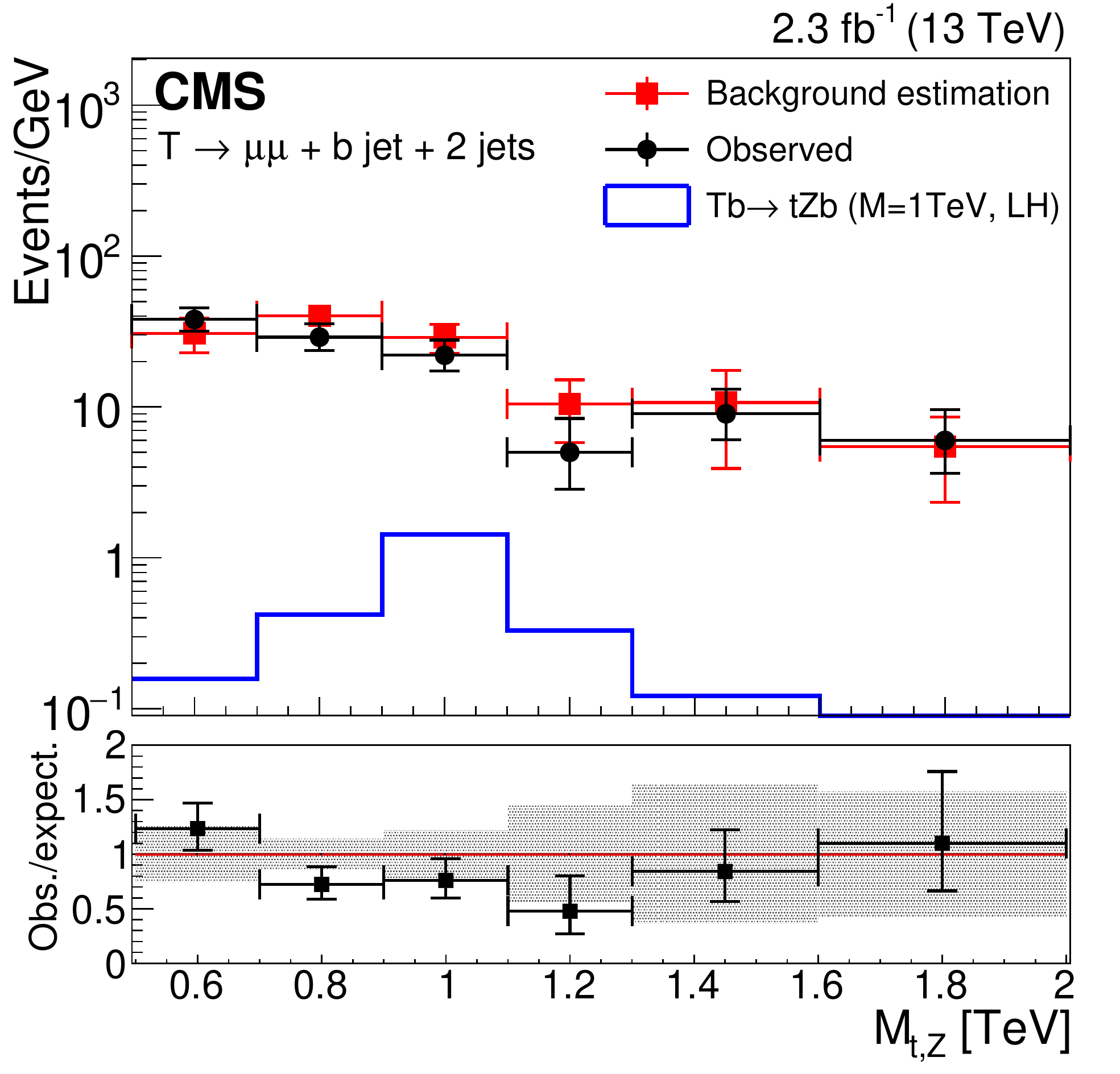}
\includegraphics[width=0.48\textwidth]{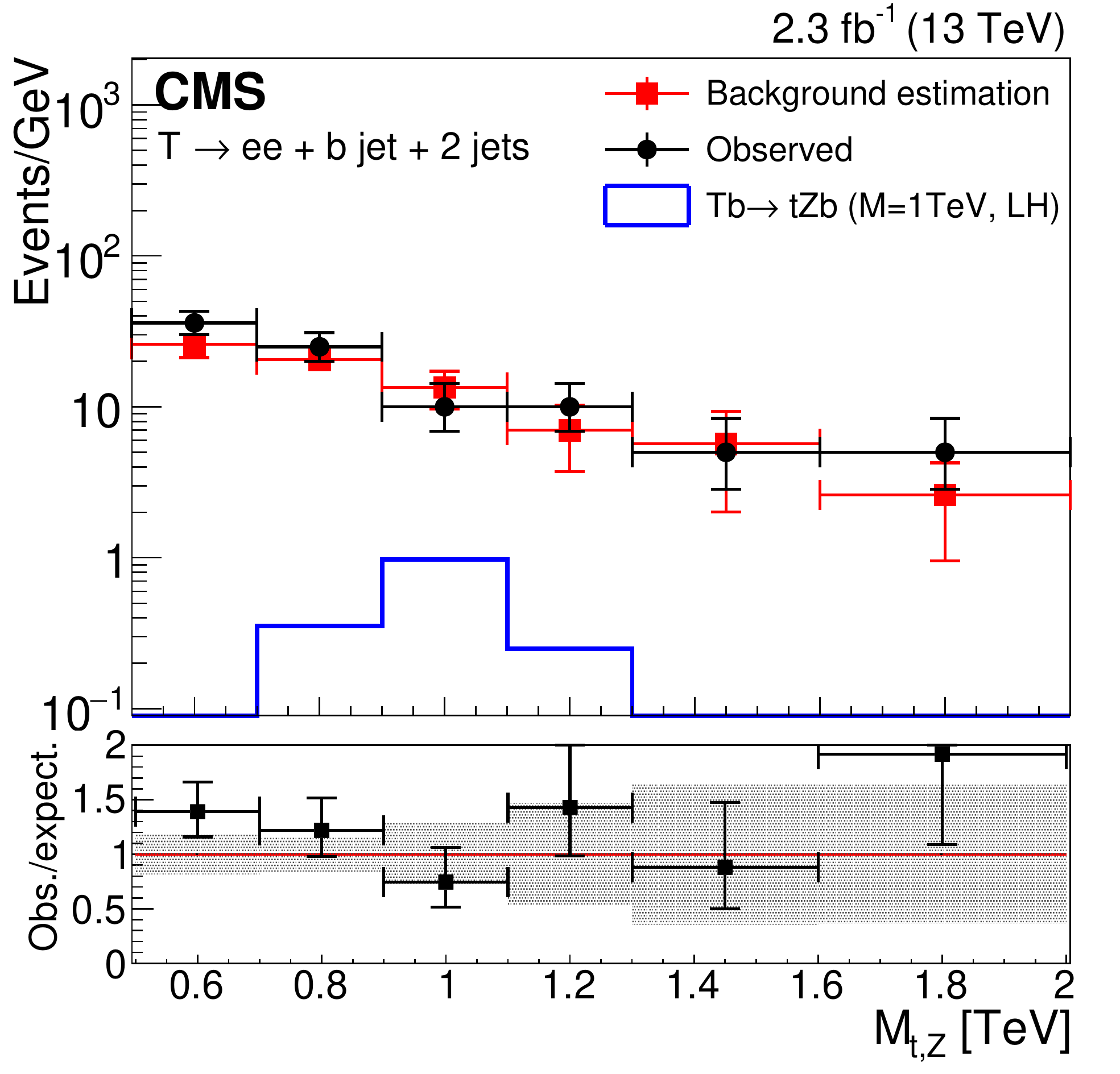}
\caption{Comparison between the background estimate and data for the T categories: fully merged region (upper-left), partially merged region (upper right), and resolved region (lower) for events with the Z boson decaying into muons (left) and electrons (right). For the fully and partially merged topologies, the sets of events with the Z boson decaying to muons and electrons are combined. For the fully merged region a shape analysis is not performed because of the small number of events, and a single bin is shown. The uncertainties in the background estimate method include both statistical and systematic components, as described in Section~\ref{sec:systematics}. The lower panel in each plot shows the ratio of the data and the background estimation, with the shaded band representing the uncertainties in the background estimate.}\label{fig:Background1}

\end{figure}

\begin{figure}[!thb]
\centering
\includegraphics[width=0.48\textwidth]{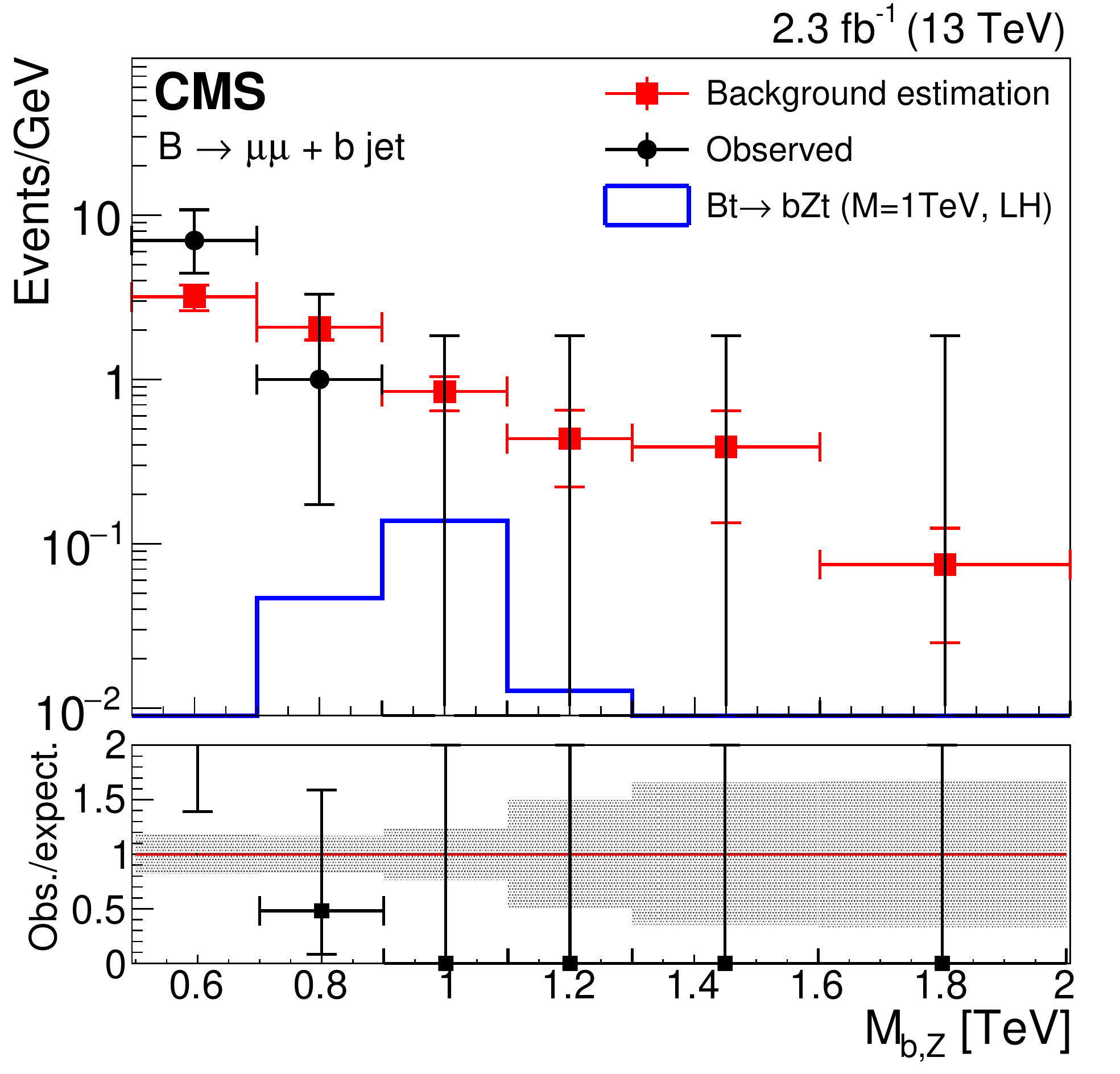}
\includegraphics[width=0.48\textwidth]{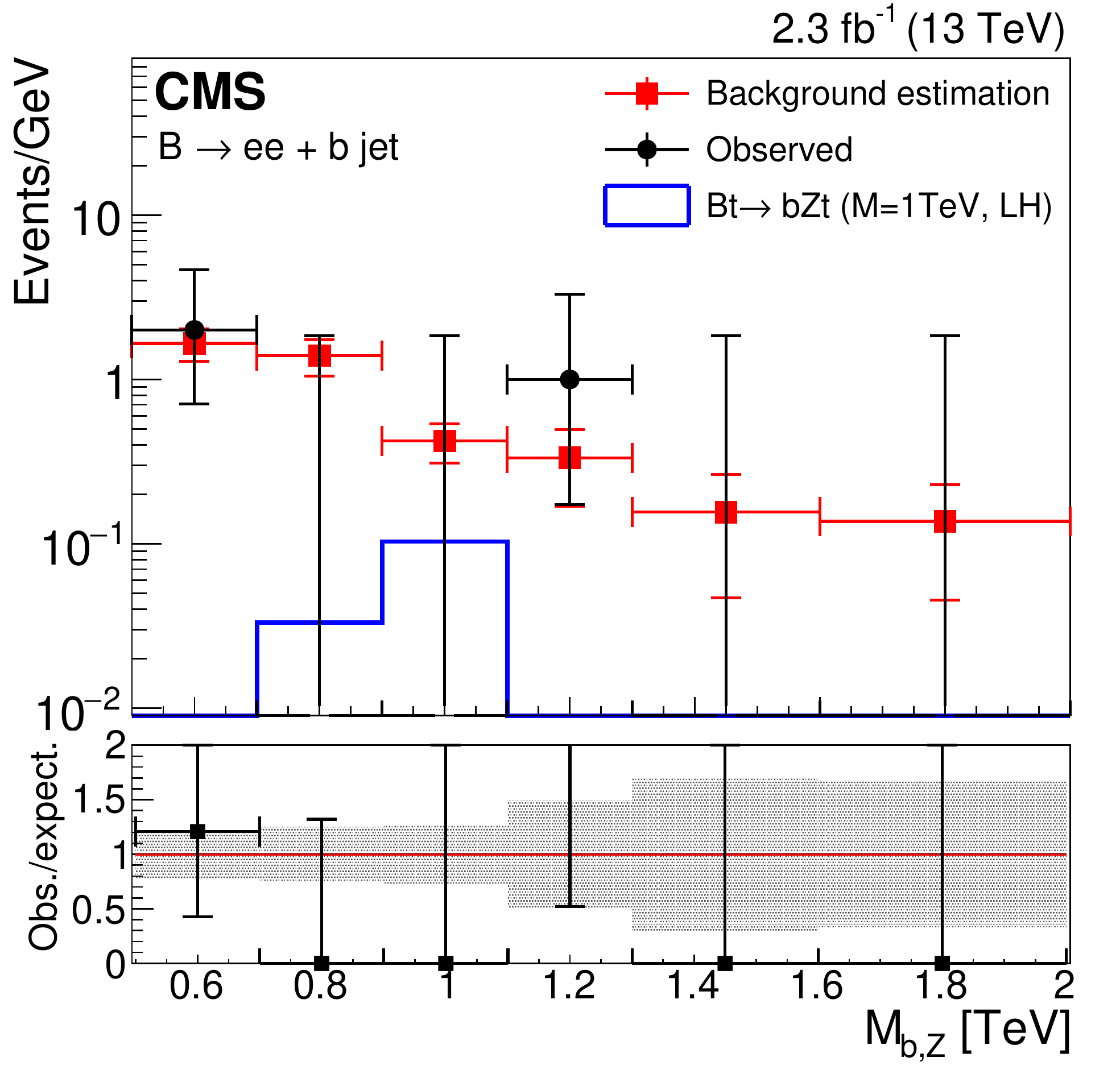}
\caption{Comparison between the background estimate and data for the B search categories: events with the Z boson decaying to muons (left) and to electrons (right). The uncertainties in the background estimate include both statistical and systematic components, as described in Section~\ref{sec:systematics}. The lower panel in each plot shows the ratio of the data and the background estimation, with the shaded band representing the uncertainties in the background estimate.}\label{fig:Background2}

\end{figure}

\section{Background estimate}\label{sec:background}
To reduce the dependence on the simulation, a background estimate primarily based on control samples in data is used. This method consists of the definition of a background-enriched control region, from which the number of events is extrapolated into the signal region. This control region is defined by the event selection described in Section~\ref{sec:selection}, but applying a b-tagged jet veto (``loose'' working point). The small signal contamination in this region has been shown not to have a significant effect on the background prediction.

The background yield in the signal region is evaluated through the following formula:
\begin{equation}\label{eq:alphaRatioMethod}
N_{\text{bkg}}(M_{\PQq,\Z}) = N_{\text{cr}}(M_{\PQq,\Z}) \,\alpha(M_{\PQq,\Z}),
\end{equation}
where $N_{\text{cr}}(M_{\PQq,\Z})$ is the number of events found in the data sample in the control region as a function of $M_{\PQq,\Z}$, $\alpha(M_{\PQq,\Z})$ is the ratio for each bin in $M_{\PQq,\Z}$ of the number of events in the signal region to the number of events in the control region, taken from simulation, and q is the t (b) jet used to reconstruct the T (B) mass.
A closure test has been performed to validate the method in a region orthogonal with respect to the signal selection. This region has been selected by requiring exactly two leptons with the same identification, \pt, and $\Delta R$ requirements as those defined for the signal region, for one or two jets and zero W and t jets. A good agreement is found between the predicted background and the observed data, supporting the validity of the method.

Comparisons between the background estimates and the observations in data are shown in Figs.~\ref{fig:Background1} and~\ref{fig:Background2}. For the fully merged topology one single bin is considered because of the small number of events in this category. The numbers of predicted background events and the number of observed events are reported in Tables~\ref{tab:Estimation1} and~\ref{tab:Estimation2} for the T and the B search respectively, together with the number of expected signal events for three example mass points. The numbers of observed events are consistent with the background predictions.

\section{Systematic uncertainties}\label{sec:systematics}
Sources of systematic uncertainty in this analysis affect both the background estimate and the signal. The effects of the systematic uncertainties on the shapes of the T and B quark reconstructed mass distributions for both signal and background processes have been investigated.

The uncertainty in the background estimate comes from a number of sources, the dominant one being the statistical uncertainties (between 12\% and 57\%, depending on the category) in the control region and the simulation. The following three systematic uncertainties are also considered. The differences between the measurements and the prediction for the closure test described previously are taken as systematic uncertainties (8--16\%). The uncertainty from the b tagging efficiency for the b, c, and light-flavor jets is evaluated by varying the b tagging scale factors (used to correct for the differences between measurements and simulation) by their uncertainties~\cite{Chatrchyan:2012jua, CMS-PAS-BTV-15-001}, giving a systematic uncertainty of between 4 and 10\%. Finally, an uncertainty (8--20\%) is included that takes into account possible mismodelling of the Z+light quark and Z+b quark fractions in simulation. This systematic uncertainty is computed by changing the Z+b fraction by 50\%~\cite{Chatrchyan:2014dha}, and re-evaluating the background through the background estimation method.

For the signal, the main sources of systematic uncertainties come from corrections that are applied to the simulation in order to match distributions in data. The scale factors for lepton identification and lepton triggers are derived from dedicated analyses, using the ``tag-and-probe'' method~\cite{Chatrchyan:2012xi, Khachatryan:2015hwa}. The uncertainties in these factors are taken as systematic uncertainties for this analysis and are found to be between 2.8 and 5.0\% for muons, between 0.4 and 1.2\% for electrons, and between 0.7 and 1.1\% for the trigger. The jet four-momenta are varied by the JES and JER uncertainties, resulting in a variation for the signal of between 0.2 and 1.9\% for the JES, and 0.1 and 2.0\% for the JER. For W and t jet tagging, the same procedure of varying the scale factors results in a systematic uncertainty of 3--8 and 18\%, respectively. The uncertainty in the b tagging efficiency is evaluated, as for the background, by scaling up and down the b tagging scale factors by their uncertainties~\cite{Chatrchyan:2012jua, CMS-PAS-BTV-15-001}, giving systematic uncertainties of between 6.0 and 13.4\%, depending on the category and on the signal mass hypothesis. Parton distribution function uncertainties are evaluated using the {NNPDF}~3.0 PDF eigenvectors~\cite{Butterworth:2015oua}. The uncertainty in the pileup simulation (0.2--2.0\%) is obtained by varying the inelastic cross section value, which controls the average pileup multiplicity, by 5\%~\cite{Aaboud:2016mmw}. Additional sources of systematic uncertainty are the integrated luminosity determination (2.7\%)~\cite{CMS-PAS-LUM-15-001} and the factorization and renormalization scales.

\section{Results}\label{sec:results}
No significant deviations from the expected background are observed in any of the search channels. We proceed with setting upper limits on the product of the production cross section and branching fraction of a T (B) quark decaying to tZ (bZ), using the predictions from the background estimation method. The 95\% confidence level (CL) exclusion limits are derived using the asymptotic CL$_\mathrm{s}$ criterion~\cite{Read:2002hq, Junk:1999kv, Cowan:2010js, ATLAS:2011tau}, with background and signal templates given by the distributions of Figs.~\ref{fig:Background1} and~\ref{fig:Background2}. Systematic uncertainties are treated as nuisance parameters.

\begin{figure}[!thb]
\centering
\includegraphics[width=0.48\textwidth]{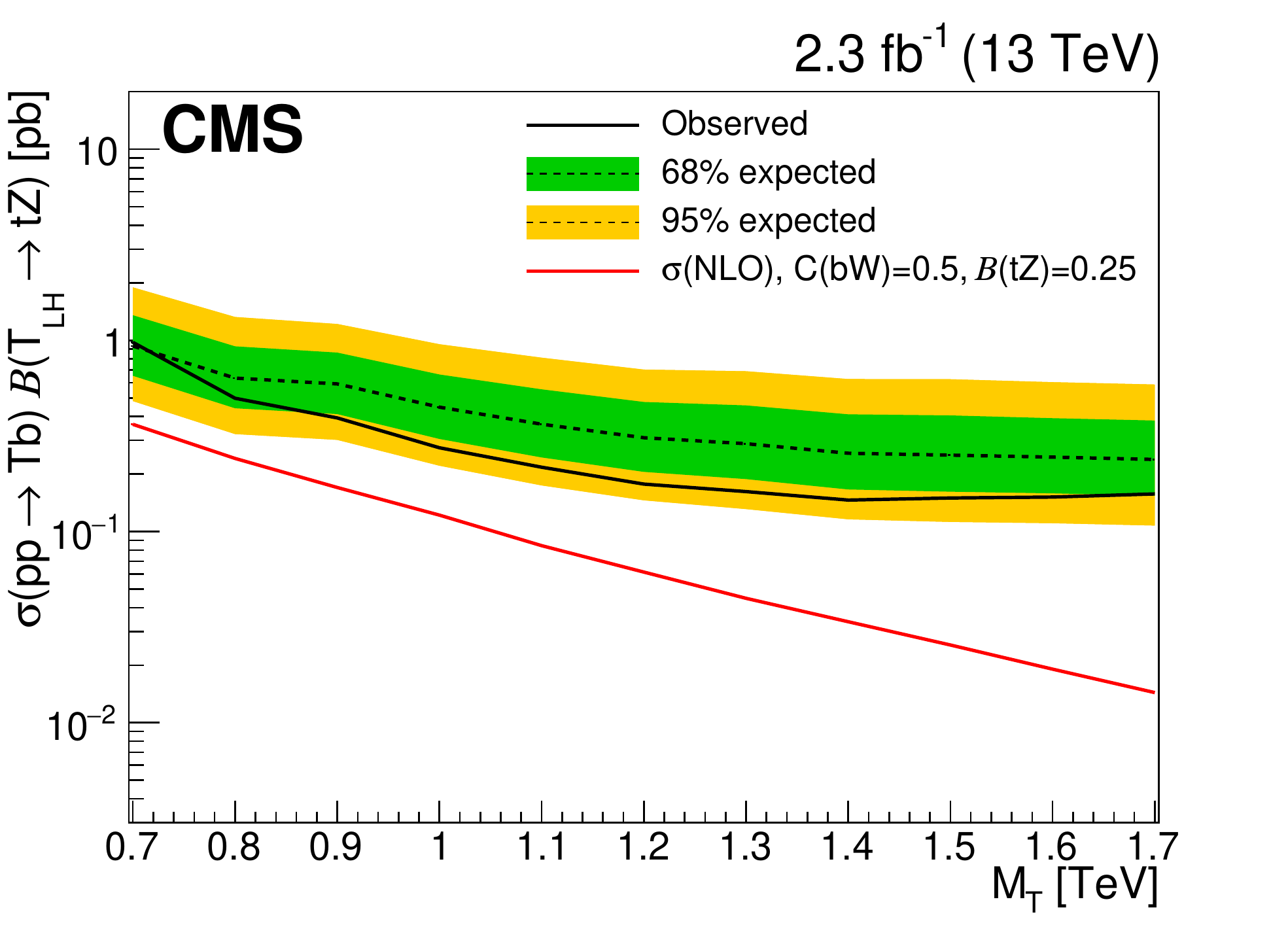}
\includegraphics[width=0.48\textwidth]{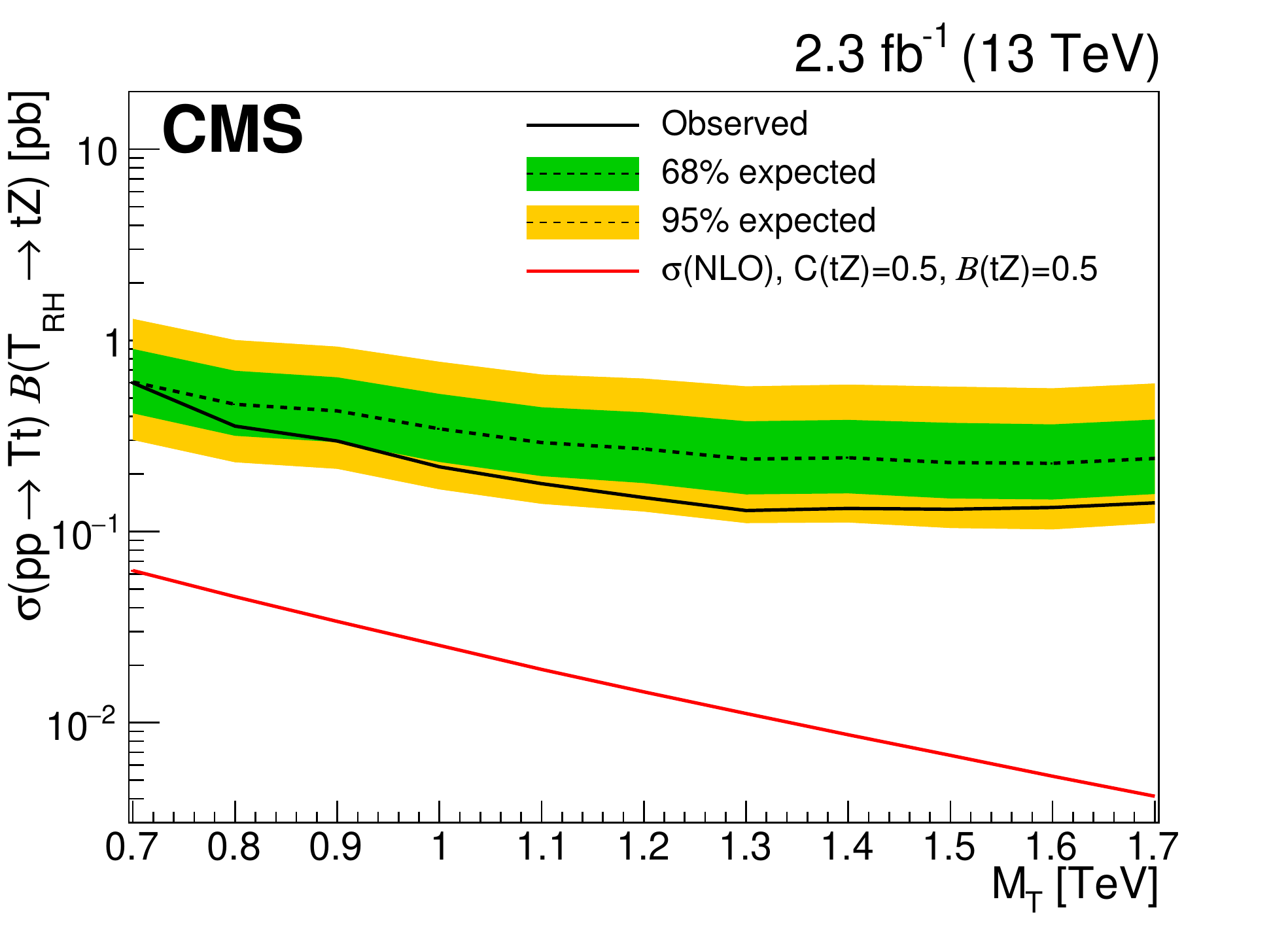}
\caption{Observed and expected 95\% CL upper limit on the product of cross section and branching fraction for the singlet LH T(b) (left) and doublet RH T(t) (right) production modes, with the T decaying to tZ. The 68\% and 95\% expected bands are shown. Theoretical cross sections as calculated at NLO in Ref.~\cite{Matsedonskyi:2014mna} are shown. The branching fraction $\mathcal{B}$(T$\to$ tZ) is 0.25 (0.5) for the left (right) plot.}\label{fig:limitTprime1}

\end{figure}

In Fig.~\ref{fig:limitTprime1}, the observed and expected limits from the four categories of the T quark search are shown combined together for the singlet LH T(b) and doublet RH T(t) production modes. Limits on the product of the cross section and branching fraction have been set, excluding values above 0.98--0.15\unit{pb} at 95\% CL, depending on the resonance mass. For an RH T(t) signal, the range between 0.60 and 0.13\unit{pb} has been excluded.

\begin{figure}[!bht]
\centering
\includegraphics[width=0.48\textwidth]{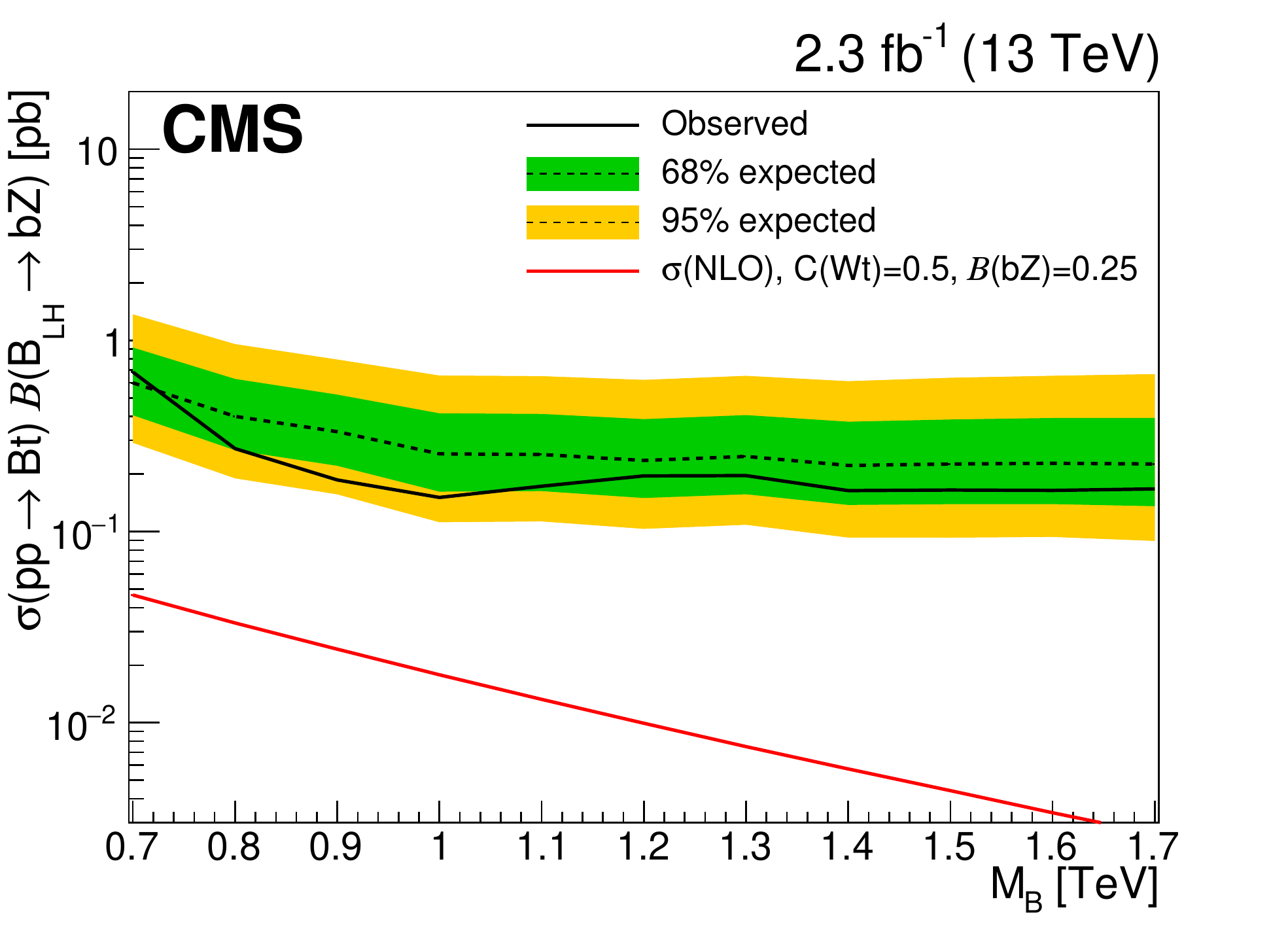}
\includegraphics[width=0.48\textwidth]{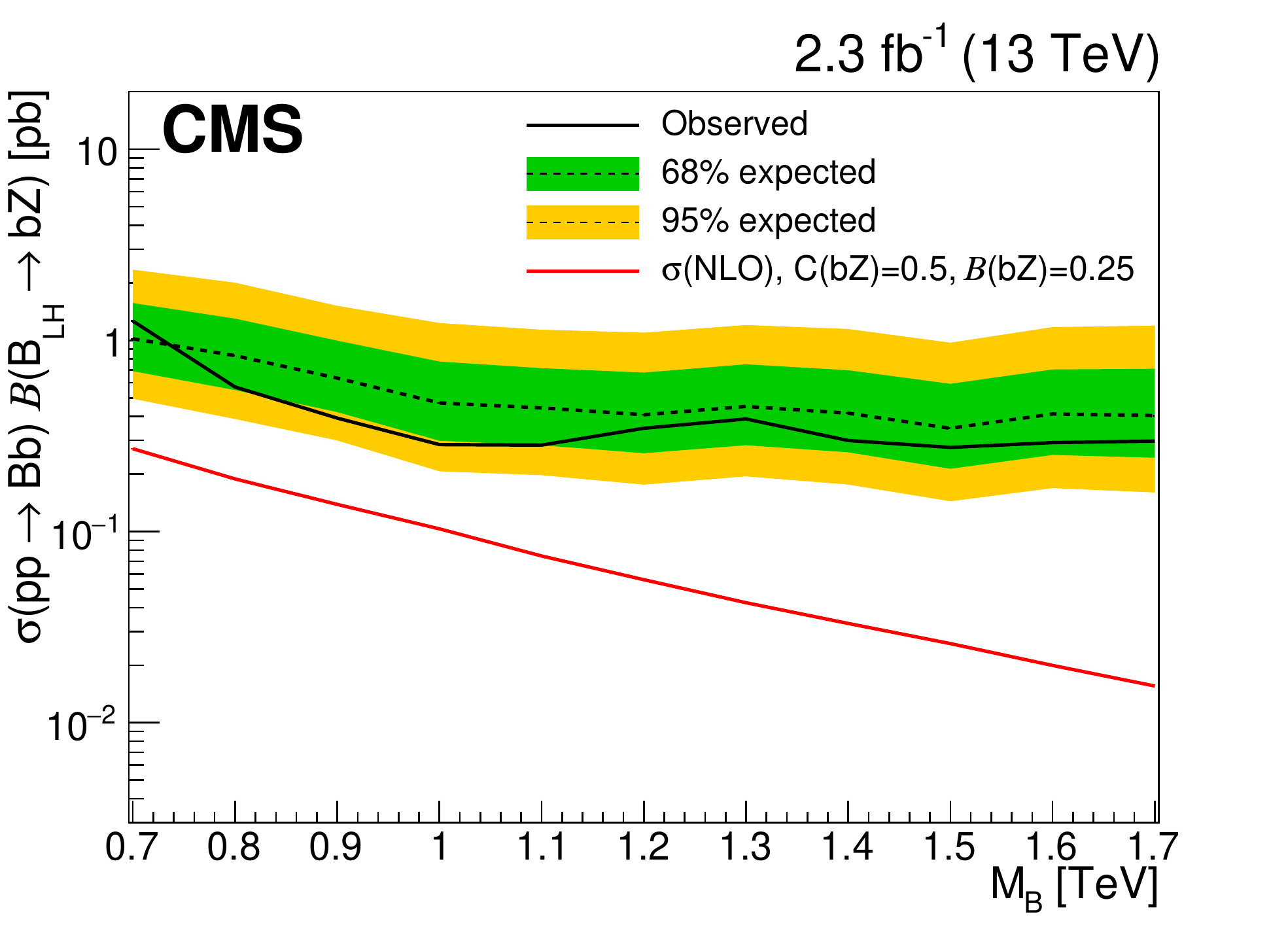}
\caption{Observed and expected 95\% CL upper limit on the product of cross section and branching fraction for the B(t) (left) and B(b) (right) signals in the singlet LH scenario, with the B decaying to bZ. The 68\% and 95\% expected bands are shown. Theoretical cross sections as calculated at NLO in Ref.~\cite{Matsedonskyi:2014mna} are shown. The branching fraction $\mathcal{B}(\PB\to \PQb\Z)$ is 0.25.}\label{fig:limitBprime}

\end{figure}

In Fig.~\ref{fig:limitBprime}, the observed and expected limits for the B quark search are shown, in cases where the B quark is produced in association with a t or a b quark for the singlet LH scenario. In this case, products of the cross section and branching fraction between 0.68 and 0.15\unit{pb} are excluded at 95\% CL in the 0.7--1.7\TeV mass range for the B(t) signal and between 1.26 and 0.28\unit{pb} in the same mass range for the B(b) signal.

Upper limits are compared with theoretical cross sections as calculated at NLO in Ref.~\cite{Matsedonskyi:2014mna}. With the present sensitivity it is not possible to exclude this particular benchmark model.

In Table~\ref{tab:Zprime}, observed and expected limits are shown for the production of a T quark via a decay of a \PZpr boson, $\PZpr\to\PQT\PQt$. The products of the cross section and branching fraction for this process are excluded between 0.31 and 0.13\unit{pb}, depending on the \PZpr mass over the range from 1.5 to 2.5\TeV and on the T mass over the range from 0.7 to 1.5\TeV. A branching fraction of 100\% is assumed for the decay of the T quark into the tZ channel. Limits for other branching fractions can be obtained by scaling the limit by the corresponding branching fraction value.

\begin{table}[!h]
\centering
\topcaption{Observed and expected 95\% CL upper limit on $\sigma \mathcal{B}$ for the $\PZpr\to\PQT\PQt$ signal. The branching fraction $\mathcal{B}(\PQT \to\PQt\Z)$ is taken to be 100\%. In order to consider different branching fractions, the limits should be scaled by the corresponding branching fraction value. The 1 and 2 standard deviation bands are given.}\label{tab:Zprime}
\begin{tabular}{cccccc}
\hline
M(\PZpr) [\TeVns{}] & M(T) [\TeVns{}] & Observed & Expected & Expected $+$ 1(2) s.d. & Expected $-$ 1(2) s.d. \\
\hline
1.5 & 0.7 & 0.31 & 0.47 & 0.71 (1.01) & 0.32 (0.23) \\
1.5 & 0.9 & 0.25 & 0.40 & 0.60 (0.85) & 0.28 (0.20) \\
1.5 & 1.2 & 0.15 & 0.26 & 0.41 (0.60) & 0.17 (0.12) \\
2.0 & 0.9 & 0.15 & 0.27 & 0.42 (0.63) & 0.18 (0.13) \\
2.0 & 1.2 & 0.13 & 0.24 & 0.37 (0.55) & 0.16 (0.11) \\
2.0 & 1.5 & 0.13 & 0.24 & 0.38 (0.57) & 0.16 (0.11) \\
2.5 & 1.2 & 0.14 & 0.24 & 0.39 (0.59) & 0.16 (0.11) \\
2.5 & 1.5 & 0.13 & 0.22 & 0.34 (0.53) & 0.14 (0.10) \\
\hline
\end{tabular}

\end{table}

\section{Summary}\label{sec:summary}
Results of a search for single production of a T quark with a charge of $+2/3$ decaying to a Z boson and a top quark and of a search for single production of a B quark with a charge of $-1/3$ decaying to a b quark and a Z boson have been presented. No deviations from the expected standard model background are observed. Limits on the product of the cross section and branching fraction for a left-handed T(b), with the T quark decaying to tZ, vary between 0.98 and 0.15\unit{pb} at 95\% confidence level and between 0.60 and 0.13\unit{pb} for a right-handed T(t) signal, for the range of resonance mass considered, which is between 0.7 and 1.7\TeV. For a left-handed B quark produced in association with a top quark and decaying to bZ, products of the cross section and branching fraction between 0.68 and 0.15\unit{pb} are excluded in the same mass range, while for a B quark produced in association with a bottom quark, products of the cross section and branching fraction between 1.26 and 0.28\unit{pb} are excluded. Additionally, products of the cross section and branching fraction for T quarks from the decay $\PZpr\to\PQT\PQt$ are excluded between 0.31 and 0.13\unit{pb}, for the range of \PZpr (T) mass considered, which is between 1.5 to 2.5 (0.7 to 1.5)\TeV. This is the first search at 13\TeV for single production of vector-like quarks in events with a Z boson decaying leptonically accompanied by boosted jets.

\begin{acknowledgments}
We congratulate our colleagues in the CERN accelerator departments for the excellent performance of the LHC and thank the technical and administrative staffs at CERN and at other CMS institutes for their contributions to the success of the CMS effort. In addition, we gratefully acknowledge the computing centres and personnel of the Worldwide LHC Computing Grid for delivering so effectively the computing infrastructure essential to our analyses. Finally, we acknowledge the enduring support for the construction and operation of the LHC and the CMS detector provided by the following funding agencies: BMWFW and FWF (Austria); FNRS and FWO (Belgium); CNPq, CAPES, FAPERJ, and FAPESP (Brazil); MES (Bulgaria); CERN; CAS, MoST, and NSFC (China); COLCIENCIAS (Colombia); MSES and CSF (Croatia); RPF (Cyprus); SENESCYT (Ecuador); MoER, ERC IUT, and ERDF (Estonia); Academy of Finland, MEC, and HIP (Finland); CEA and CNRS/IN2P3 (France); BMBF, DFG, and HGF (Germany); GSRT (Greece); OTKA and NIH (Hungary); DAE and DST (India); IPM (Iran); SFI (Ireland); INFN (Italy); MSIP and NRF (Republic of Korea); LAS (Lithuania); MOE and UM (Malaysia); BUAP, CINVESTAV, CONACYT, LNS, SEP, and UASLP-FAI (Mexico); MBIE (New Zealand); PAEC (Pakistan); MSHE and NSC (Poland); FCT (Portugal); JINR (Dubna); MON, RosAtom, RAS, RFBR and RAEP (Russia); MESTD (Serbia); SEIDI and CPAN (Spain); Swiss Funding Agencies (Switzerland); MST (Taipei); ThEPCenter, IPST, STAR, and NSTDA (Thailand); TUBITAK and TAEK (Turkey); NASU and SFFR (Ukraine); STFC (United Kingdom); DOE and NSF (USA).
\end{acknowledgments}

\bibliography{auto_generated}

\cleardoublepage \appendix\section{The CMS Collaboration \label{app:collab}}\begin{sloppypar}\hyphenpenalty=5000\widowpenalty=500\clubpenalty=5000\textbf{Yerevan Physics Institute,  Yerevan,  Armenia}\\*[0pt]
A.M.~Sirunyan, A.~Tumasyan
\vskip\cmsinstskip
\textbf{Institut f\"{u}r Hochenergiephysik,  Wien,  Austria}\\*[0pt]
W.~Adam, E.~Asilar, T.~Bergauer, J.~Brandstetter, E.~Brondolin, M.~Dragicevic, J.~Er\"{o}, M.~Flechl, M.~Friedl, R.~Fr\"{u}hwirth\cmsAuthorMark{1}, V.M.~Ghete, C.~Hartl, N.~H\"{o}rmann, J.~Hrubec, M.~Jeitler\cmsAuthorMark{1}, A.~K\"{o}nig, I.~Kr\"{a}tschmer, D.~Liko, T.~Matsushita, I.~Mikulec, D.~Rabady, N.~Rad, B.~Rahbaran, H.~Rohringer, J.~Schieck\cmsAuthorMark{1}, J.~Strauss, W.~Waltenberger, C.-E.~Wulz\cmsAuthorMark{1}
% ...
\vskip\cmsinstskip
% ...
1:~~Also at Vienna University of Technology, Vienna, Austria\\
% ...

\end{sloppypar}
\end{document}